\documentclass[article,pre,aps,groupaddress,showpacs,prd,onecolumn]{revtex4-1}
\usepackage{indentfirst}
\usepackage{bm}
\usepackage{epsfig}
\usepackage{float}
\usepackage{graphicx}
\usepackage{epstopdf}
\usepackage{epsfig,amsmath,graphicx,amssymb,overpic}
\usepackage{graphicx}
\usepackage{dcolumn}
\usepackage{bm}
\usepackage{graphicx}
\usepackage{subfigure}
\usepackage{diagbox}
\usepackage{booktabs}
\usepackage{amsmath}
\usepackage{multirow}
\usepackage{threeparttable}

\def\be{\begin{equation}}
\def\ee{\end{equation}}
\def\bee{\begin{eqnarray}}
\def\ene{\end{eqnarray}}
\def\bes{\begin{subequations}}
\def\ees{\end{subequations}}
\makeatletter 
\renewcommand{\section}{\@startsection{subsection}{1}{0mm} {-\baselineskip}{0.5\baselineskip}{\bf\leftline}} \makeatother

\makeatletter 
\renewcommand{\subsection}{\@startsection{subsection}{1}{0mm} {-\baselineskip}{0.5\baselineskip}{\bf\leftline}} \makeatother

\begin{document}

\begin{center}
\LARGE\bf Spin-current-controlled anisotropic deformation of magnetic lump solitons
\end{center}

\begin{center}
\rm Xiao-Qi Cui$^{\textrm{a,b}}$, Xin-Wei Jin$^{\textrm{d},\ast}$, Shiyi Wang$^{\textrm{a,b}}$, Xiao-Yong Wen$^{\textrm{e}}$, Zhan-Ying Yang$^{\textrm{a,b,c},\dagger}$
\end{center}

\begin{center}
\begin{footnotesize} \sl
$^{\textrm{a}}$School of Physics, Northwest University, Xi'an 710127, China\\
$^{\textrm{b}}$Peng Huanwu Center for Fundamental Theory, Xi'an 710127, China\\
$^{\textrm{c}}$Shaanxi Key Laboratory for Theoretical Physics Frontiers, Xi'an 710127, China\\
$^{\textrm{d}}$Department of Physics, Zhejiang Normal University, Jinhua 321004, China\\
$^{\textrm{e}}$School of Applied Science, Beijing Information Science and Technology University, Beijing 100192, China\\
$\ast$Corresponding author at: School of Physics, Northwest University, Xi'an 710127, China.\\
                                E-mail address: jinxinwei@zjnu.edu.cn.\\
                                E-mail address: zyyang@nwu.edu.cn.\\
\end{footnotesize}
\end{center}

\vspace*{2mm}

\begin{center}
\begin{minipage}{15.5cm}
{\bf ABSTRACT}
\parindent 20pt  \small 
We investigate a (2+1)-dimensional nonlinear spin system containing an effective spin-current transport term. Based on its integrable structure, exact magnetic lump solutions are constructed on a rotating spin background, including both fundamental and higher-order configurations generated via the Darboux transformation. The obtained excitations are doubly localized in spatial directions, while their temporal evolution is characterized by intrinsic spin precession rather than translational motion of the localized envelope. It is shown that the effective spin-current contribution enters the localization coordinate and acts as a geometric control parameter for the spatial structure of the solutions. In particular, spin current induces anisotropic deformation of the localized profile, leading to a continuous transition toward a quasi-one-dimensional soliton-like state under specific parameter regimes. More importantly, this deformation mechanism is found to be universal across different hierarchical lump structures, including both fundamental and higher-order solutions, indicating that spin current governs a unified structural modulation law for the entire family of localized spin excitations. These results provide an analytically tractable example of spin-current-controlled anisotropic deformation and dimensional crossover in nonlinear spin systems, and further reveal a universal mechanism for geometric control of localized spin textures beyond individual solution types.
\end{minipage}
\end{center}

\begin{center}
\begin{minipage}{15.5cm}
\end{minipage}
\end{center}
\newpage
\section*{1. Introduction}

Spin current can transfer angular momentum to local magnetic moments and thereby manipulate localized magnetic structures~\cite{001,002,003,004}. As a result, current-driven manipulation of domain walls, magnetic solitons, and skyrmions has attracted considerable attention, with previous studies mainly focusing on their motion, deformation, and stability~\cite{005,006,007,008,009}. These studies indicate that spin current provides an effective approach for manipulating localized magnetic structures and is therefore of considerable importance for magnetic information transport. To date, investigations of spin-current-driven effects have primarily concentrated on quasi-one-dimensional localized excitations, where spin current is mainly reflected in the modulation of transport behavior and dynamical properties~\cite{010,011,012}. Most existing studies emphasize translational dynamics, drift velocity, depinning, or the stability of localized objects, whereas much less attention has been paid to how an effective spin-current term influences the internal localization geometry of two-dimensional non-topological localized excitations. Compared with quasi-one-dimensional excitations, two-dimensional localized structures possess additional spatial degrees of freedom and may therefore exhibit richer dynamical behaviors. Consequently, investigating spin-current-driven spatially double-localized excitations in nonlinear magnetic systems is of considerable interest~\cite{013,014}.

Lump solitons are rationally localized wave structures in two spatial dimensions and have been widely studied in integrable nonlinear systems~\cite{015,016,017,018}. As spatially double-localized excitations with two-dimensional degrees of freedom, they may exhibit richer structural evolution under external driving~\cite{019,020}. In recent years, lump structures have been obtained analytically in the Kadomtsev-Petviashvili equation and other integrable models~\cite{021,022,023,024,025,026,027}. Although exact lump solutions have been reported in various integrable systems, the controllable deformation of their localization geometry by a current-like term remains insufficiently explored~\cite{028,029,030,031,A1}. It is worth noting that the recent first experimental observation of lump solitons in nonlinear optical systems demonstrated that such spatially double-localized structures are not merely theoretical solutions, but can also exist as real physical excitations in nonlinear media~\cite{032}. This suggests that studies of lump solitons are gradually extending from the construction of analytical solutions to the investigation of their dynamical evolution and controllable modulation in physical systems. Against this background, a natural and interesting question arises: whether an effective spin-current parameter can enter the localization coordinates of a two-dimensional magnetic lump and thereby induce structural transitions between different localized states. However, this issue has received relatively little attention.

In this work, we investigate a (2+1)-dimensional nonlinear spin system containing an effective spin-current transport term. Based on its integrable structure, we construct an exact magnetic lump solution on a rotating spin background and analyze its intrinsic spin-precession dynamics. The results show that spin current not only modifies the spatial structure of lump excitations, but, more importantly, the effective spin-current parameter enters the localization coordinate of the lump, thereby serving as an anisotropic control parameter for its spatial confinement. These results further deepen the understanding of spin-current modulation mechanisms of localized structures and provide a new theoretical perspective for understanding structural evolution and state transitions of complex localized excitations. The remainder of this paper is organized as follows. In Section 2, we introduce the nonlinear spin model and establish the analytical framework for constructing exact magnetic lump solutions. In Section 3, we investigate the fundamental properties, spatial morphology, and spin-precession dynamics of the lump, followed by its spin-current-controlled structural modulation and transitions between different localized states. In Section 4, the spin-current-induced structural modulation is further extended to higher-order magnetic lumps, where aggregation-separation dynamics and the universality of the modulation mechanism are investigated. Finally, Section 5 concludes the paper with a summary and discussion.

\section*{2. Nonlinear spin model and magnetic lump structures}

We consider an effective (2+1)-dimensional nonlinear spin model with a current-like transport contribution~\cite{033,034,035,036}
\begin{equation}
\begin{aligned}
\vec{S}_t &= (\vec{S}\times \vec{S}_y + u\vec{S})_x + \nu_1 \vec{S}_x, \\
u_x &= -\vec{S}\cdot(\vec{S}_x \times \vec{S}_y),
\end{aligned}
\label{hfe}
\end{equation}
where $\vec{S}=(S^x,S^y,S^z)$ denotes the spin vector satisfying the constraint $|\vec{S}|=1$, and $u(x,y,t)$ is a scalar potential determined self-consistently by the spin configuration. The dynamics of Eq.~(\ref{hfe}) contains several coupled nonlinear mechanisms. The term $(\vec{S}\times \vec{S}_y)_x$ characterizes the nonlinear rotational interaction generated by spatial variations of the spin field, while the scalar field $u$ is induced through $u_x=-\vec{S}\cdot(\vec{S}_x \times \vec{S}_y)$, which reflects the local twisting structure and geometric chirality of the two-dimensional spin texture. The parameter $\nu_1$ is a dimensionless coefficient representing an effective spin-current-like convective contribution along the $x$-direction.

To construct localized excitations of Eq.~(\ref{hfe}), we consider the following linear spectral problem~\cite{033}
\begin{equation}
\begin{aligned}
\Phi_x &= \Gamma \Phi
=\frac{\mathrm{i}}{2}\lambda \mathrm{S}\Phi, \\
\Phi_t &= \lambda \Phi_y+\Xi\Phi
=\lambda\Phi_y+\frac{\lambda}{4}
\left(
[\mathrm{S},\mathrm{S}_y]
+2\mathrm{i}(\nu_1+u)\mathrm{S}
\right)\Phi,
\end{aligned}
\label{lax}
\end{equation}
where
\begin{equation*}
\mathrm{S}=
\begin{pmatrix}
S^z & S^- \\
S^+ & -S^z
\end{pmatrix},
\qquad
S^\pm=S^x\pm \mathrm{i}S^y.
\end{equation*}
The matrix $\mathrm{S}$ can equivalently be expressed as $\mathrm{S}=S^x\sigma_1+S^y\sigma_2+S^z\sigma_3$, where $\sigma_1$, $\sigma_2$, and $\sigma_3$ denote the Pauli matrices. The spectral parameter $\lambda$ satisfies the nonisospectral conditions $\lambda_t-\lambda \lambda_y=0$, $\lambda_x=0$. The compatibility condition of Eq.~(\ref{lax}) reproduces the nonlinear spin system~(\ref{hfe}), thereby establishing the integrable structure of the model and providing the basis for constructing localized excitations through the Darboux transformation (DT)~\cite{018,037,038,039}.

To generate localized spin structures, we introduce the rotating spin background $\vec{S}_0 =(a\cos\Theta,\,-a\sin\Theta,\,b)$, where $\Theta = kx + my + (bm + c + \nu_1)kt$ and $u_0=c$, subject to the constraint $a^2+b^2=1$. This seed solution represents a helical spin configuration with a space-time dependent phase. Substituting the background into Eq.~\eqref{lax}, the corresponding eigenfunctions are
\begin{equation}
\Phi=
\begin{pmatrix}
\mathcal{F}\\[2mm]
\mathcal{G}
\end{pmatrix}
=
\begin{pmatrix}
C_{1}F_{1}\!\left(\dfrac{\lambda t+y}{\lambda}\right)
e^{\mathcal{A}+\mathcal{B}}
+
C_{2}F_{2}\!\left(\dfrac{\lambda t+y}{\lambda}\right)
e^{-\mathcal{A}+\mathcal{B}}
\\[4mm]
\dfrac{
C_{1}\mathcal{M}_{1}
F_{1}\!\left(\dfrac{\lambda t+y}{\lambda}\right)
e^{\mathcal{A}+\mathcal{B}}
+
C_{2}\mathcal{M}_{2}
F_{2}\!\left(\dfrac{\lambda t+y}{\lambda}\right)
e^{-\mathcal{A}+\mathcal{B}}
}
{\mathcal{J}}
\end{pmatrix},
\label{laxj1}
\end{equation}
where
\begin{equation*}
\begin{aligned}
\alpha & = bm+c+\nu_{1}, \quad \beta = \sqrt{(2bk\lambda-k^{2}-\lambda^{2})\alpha^{2}}, 
\quad
\Theta = kx+my+\alpha kt, \quad \mathcal{J} = a\lambda\alpha e^{\mathrm{i}\Theta},
\\[1mm]
\delta(\varepsilon) & =\sum\limits_{j=0}^{N}(e_j+\textrm{i}d_j)\varepsilon^{2j},\ (j=0, 1,2... ),
\quad
\mathcal{A} =\dfrac{\left[\alpha y-\lambda x+\delta(\varepsilon)\right]\beta}
{2\lambda\alpha},
\quad
\mathcal{B}  =-\dfrac{\mathrm{i}\left[(k\alpha-\lambda m)y-k\lambda x\right]}{2\lambda},
\\[1mm]
\mathcal{M}_{1} & =\mathrm{i}\beta-(b\lambda-k)\alpha,
\quad
\mathcal{M}_{2}=-\mathrm{i}\beta-(b\lambda-k)\alpha, \quad C_1=-C_2=\tfrac{1}{\varepsilon}.
\end{aligned}
\end{equation*}
Within the generalized DT framework, doubly localized magnetic lump solutions can be systematically constructed. The detailed derivation of the iterative generalized $(r,N-r)$-fold DT is presented in Appendix A. Starting from the eigenfunctions in Eq.~\eqref{laxj1}, we apply the first-order DT \eqref{xjst} to the rotating background and take the rational localization limit, yielding the following magnetic lump solution for $a=k=1$ and $b=m=0$
\begin{equation}
\begin{aligned}
S^x &=\frac{P\cos\theta+4x(\rho^2-1)\sin\theta}{(\rho^2+1)^2},
\quad
S^y=\frac{-P\sin\theta+4x(\rho^2-1)\cos\theta}{(\rho^2+1)^2},
\\[2pt]
S^z &=-\frac{8(c+\nu_1)x\widetilde{y}}{(\rho^2+1)^2},
\quad
u=\frac{R}{(\rho^2+1)^2},
\end{aligned}
\label{hf1s}
\end{equation}
with
\begin{equation*}
\begin{aligned}
\rho^2 & =x^2+(c+\nu_1)^2\widetilde{y}^2,
\quad
\theta=x+(c+\nu_1)t,
\quad
\widetilde{y}= y-\tfrac{1}{c+\nu_1}
\\[1mm]
P & =8x^2-(\rho^2+1)^2,\quad R =c\rho^4-(2c+4\nu_1)\rho^2+8(c+\nu_1)x^2-3c-4\nu_1.
\end{aligned}
\label{rho}
\end{equation*}
The obtained solution exhibits strong localization in both spatial directions, forming a typical lump-type spin excitation. Owing to the phase dependence inherited from the rotating background, the spin components possess an intrinsic time-dependent precession, whereas the overall localization profile remains spatially confined.

Moreover, the transport term $\nu_1 \vec{S}_x$ introduces an additional tunable parameter into the system, which can influence both the dynamical evolution and geometric characteristics of the localized structures. The explicit dependence of $\rho^2$ on $\nu_1$ indicates that the current-like parameter can modify the anisotropic localization scale of the lump. This point will be analyzed quantitatively in the next section.

\section*{3. Effective-current-controlled anisotropic localization of magnetic lumps}

In this section, we analyze the spatial localization and internal spin dynamics of the magnetic lump solution. We first use the case $\nu_1=0$ as a reference state to distinguish the stationary localized envelope from the intrinsic spin precession of the background. We then examine how the effective current-like parameter $\nu_1$ enters the localization coordinate and quantitatively controls the anisotropic confinement of the lump. Finally, the singular limiting case $\nu_1=-c$ is discussed, where the transverse localization length diverges and the two-dimensional lump approaches a line-localized soliton-like state.

\subsection*{3.1 Fundamental properties and spin-precession dynamics of magnetic lump states}

We first consider the current-free \(\nu_1=0\). In this case, based on Eq.~\eqref{hf1s} and $c=1$, the system admits a spatially localized spin configuration characterized by the phase variable \(\theta=x+t\) and the radial localization variable $\rho^2=x^2+(y-1)^2$, which together determine the spatial-temporal structure of the solution. At this stage, both spatial localization and temporal dependence are embedded in the spin configuration, but their physical roles will be clarified through the following analysis.

\begin{figure}[!htbp]
	\centering
	\includegraphics[scale=.38]{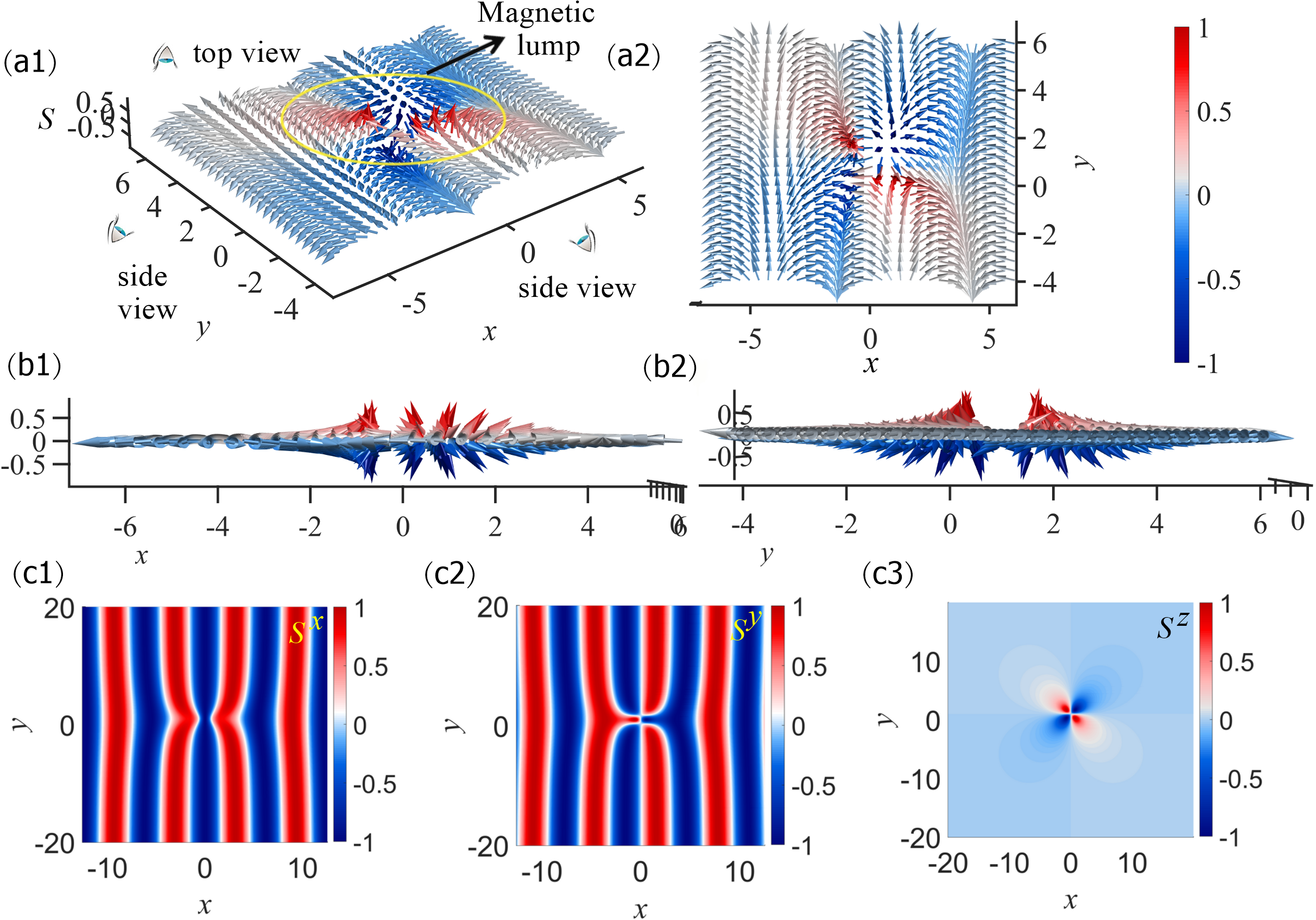}
	\caption{(Color online) Spatial spin texture of the fundamental magnetic lump at $t=0$ and $\nu_1=0$. (a1) Three-dimensional spin texture of the magnetic lump represented by the spin vector $\vec{S}=(S^x,S^y,S^z)$. (a2) Top view of the three-dimensional spin texture of the magnetic lump. (b1)-(b2) Side views along the $x$- and $y$-directions, illustrating their doubly localized spatial morphology. (c1)-(c3) Spatial distributions of the three spin components $S^x$, $S^y$, and $S^z$, respectively. The stripe-like patterns in $S^x$ and $S^y$ originate from the rotating spin background, while their local bending around the core reflects the localized lump deformation. The $S^z$ component provides a clearer signature of the localized core, exhibiting a four-petal structure. }\label{fig1}
\end{figure}

The spatial structure of the magnetic lump state at \(t=0\) is shown in Fig.~\ref{fig1}(a1)-(b2). It can be observed that the spin configuration exhibits a strongly localized structure centered in the two-dimensional plane. A pronounced twisting region appears in the core of the excitation, while the surrounding region gradually approaches the background configuration. The three-dimensional spin distribution in Fig.~\ref{fig1}(a1) provides a global view of the localized lump embedded in the spin environment, whereas the projected views in Fig.~\ref{fig1}(a2)-(b2) further confirm that the excitation is confined in both spatial directions, forming a doubly localized structure.

To further characterize the internal spatial composition of the excitation, the spin components are presented in Fig.~\ref{fig1}(c1)-(c3). The stripe-like patterns in $S^x$ and $S^y$ mainly originate from the rotating background, while their local bending near the core reflects the rational lump deformation. The out-of-plane component $S^z$ provides a clearer indicator of the localized core. At this stage, these results only describe the spatial distribution of each component at a fixed time, without explicitly addressing the role of temporal evolution. The different spatial patterns of the spin components provide a basis for understanding the subsequent dynamical behavior of the system.

\begin{figure}[!htbp]
	\centering
	\includegraphics[scale=.24]{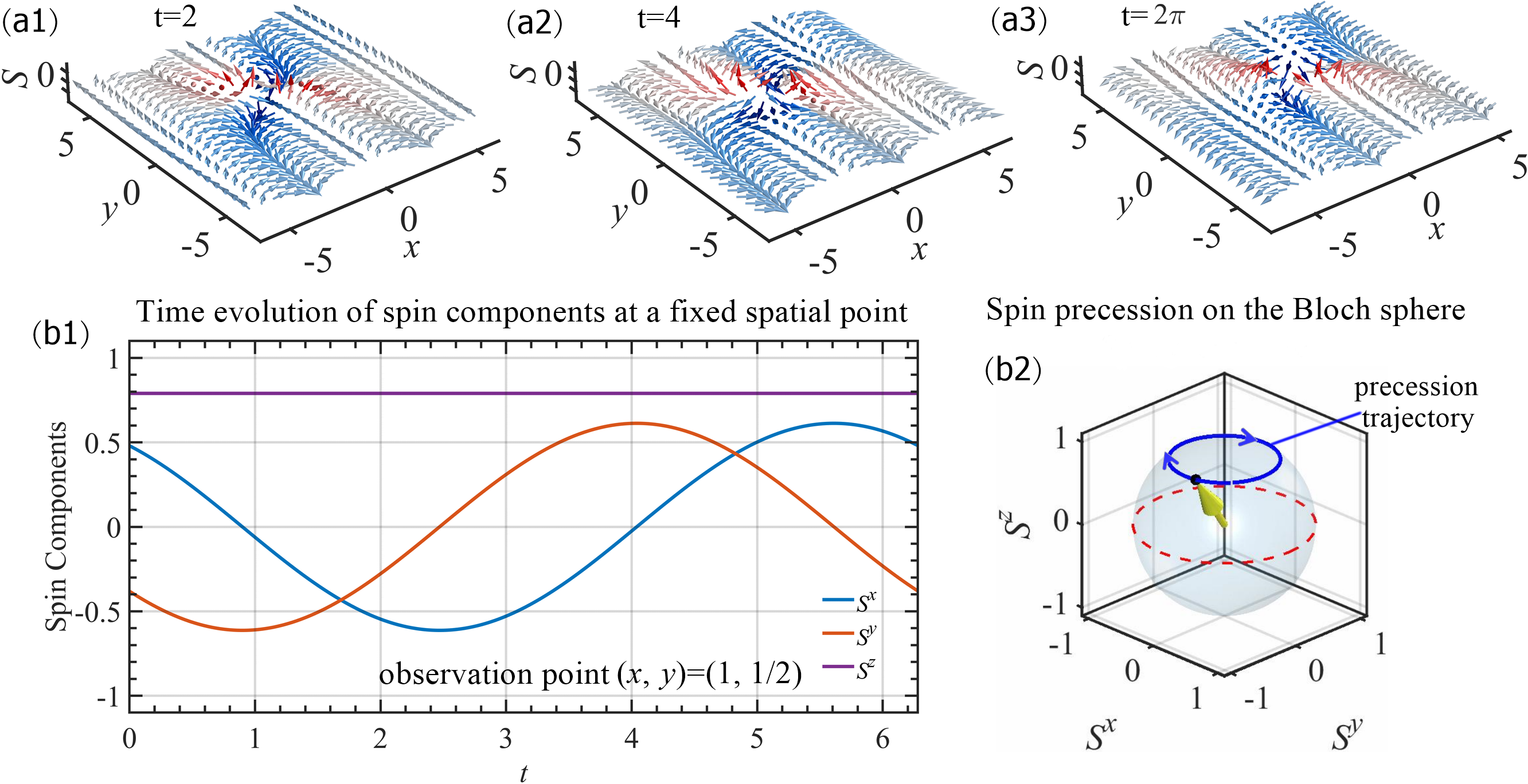}
	\caption{(Color online) Dynamical evolution and spin-precession behavior of the magnetic lump state at \(\nu_1=0\). 
(a1)-(a3) Time evolution of the spin configuration, showing that the lump remains spatially localized while the surrounding background undergoes continuous rotation. 
(b1) Time evolution of \(S^x\), \(S^y\), and \(S^z\) at a fixed spatial point, where \(S^x\) and \(S^y\) exhibit periodic oscillations characteristic of spin precession. 
(b2) Corresponding Bloch-sphere trajectory of the local spin vector, forming a closed orbit that directly reflects steady spin-precession dynamics. }\label{fig2}
\end{figure}

To clarify the role of time evolution, the spin configuration at different time instances is further examined in Fig.~\ref{fig2}(a1)-(a3). It is observed that although the spin texture changes with time, the spatial localization profile remains unchanged. This indicates that the magnetic lump does not exhibit translational motion, and the evolution of the system is not associated with the displacement of the localized structure. Instead, the apparent time dependence originates from the evolution of the spin orientation on a fixed spatial envelope. More specifically, the radial localization variable $\rho(x,y)$ is independent of time, while the temporal dependence enters only through the phase variable \(\theta=x+t\), which governs the rotation of the spin orientation. As a result, the magnetic lump remains spatially stationary, while the surrounding spin background undergoes continuous rotation.

To further understand the intrinsic dynamical behavior of the system, the temporal evolution of the spin components at a fixed spatial point is shown in Fig.~\ref{fig2}(b1). It is found that the in-plane components \(S^x\) and \(S^y\) exhibit periodic oscillations, whereas the out-of-plane component \(S^z\) remains nearly unchanged during the evolution. This behavior indicates that the local spin vector undergoes a periodic rotation in the spin plane, corresponding to a precession motion around the \(z\)-axis. These results indicate that the in-plane components ($S^x$ and $S^y$) primarily describe the rotational modulation of the background, whereas the out-of-plane component ($S^z$) determines the intrinsic localized core structure of the excitation, thereby establishing a clear relationship between the dynamical evolution and the spatially localized structure. The corresponding trajectory on the Bloch sphere is shown in Fig.~\ref{fig2}(b2), where the closed orbit directly demonstrates the periodic nature of the local spin motion. From a geometric perspective, the spin vector evolves along a closed loop while preserving its magnitude, confirming the existence of intrinsic spin-precession dynamics in the system.

These results indicate that the obtained magnetic lump corresponds to a spatially stationary two-dimensional spin structure embedded in a rotating spin background, possessing intrinsic spin-precession dynamics. It can therefore be regarded as a ``static two-dimensional magnetic lump structure with internal spin precession'', which provides a fundamental example of localized spin textures governed by nonlinear spin dynamics.

\subsection*{3.2 Spin-current-controlled structural modulation}

We next investigate the modulation effect of spin current on the spatial morphology of magnetic lump structures. 
Unlike conventional spin-current-driven magnetic excitations, where spin current mainly induces translational motion or transport behavior, the present system exhibits a fundamentally different modulation behavior. 
From Eq.~(\ref{hf1s}) and $c=1$, it can be observed that the spin-current parameter $\nu_1$ directly enters the localization coordinate through
\begin{equation}
\rho^2 =x^2+(1+\nu_1)^2\widetilde{y}^2,\quad \widetilde{y}= y-\frac{1}{1+\nu_1},\quad Y=(1+\nu_1)\widetilde{y}.
\label{rho}
\end{equation}
Therefore, the spin current does not merely influence the dynamical evolution of the excitation, but instead directly regulates the spatial localization structure of the magnetic lump.

Eq.~\eqref{rho} further shows that the modulation induced by spin current is strongly anisotropic. 
In particular, the spin-current parameter mainly affects the localization behavior along the $y$-direction, whereas the localization along the $x$-direction remains comparatively stable. 
As a consequence, the spatial morphology of the magnetic lump continuously evolves under spin-current driving, leading to deformation of the localized structure and redistribution of localized spin intensity.

To quantitatively characterize the spin-current-induced expansion along the $y$-direction, we further consider the cross-sectional profile of $S^z$ at a fixed spatial position $x=x_0$. According to the coordinates $Y=(\nu_1+1)y-1$, the anisotropy ratio can be defined as $\eta=\frac{W_x}{W_y}=\frac{1}{\left|1+\nu_1\right|}$. This result indicates that the localization width continuously increases as $\nu_1$ approaches $-1$, implying progressive weakening of spatial confinement along the $y$-direction under spin-current modulation.

\begin{figure}[!htbp]
\centering
\includegraphics[scale=.22]{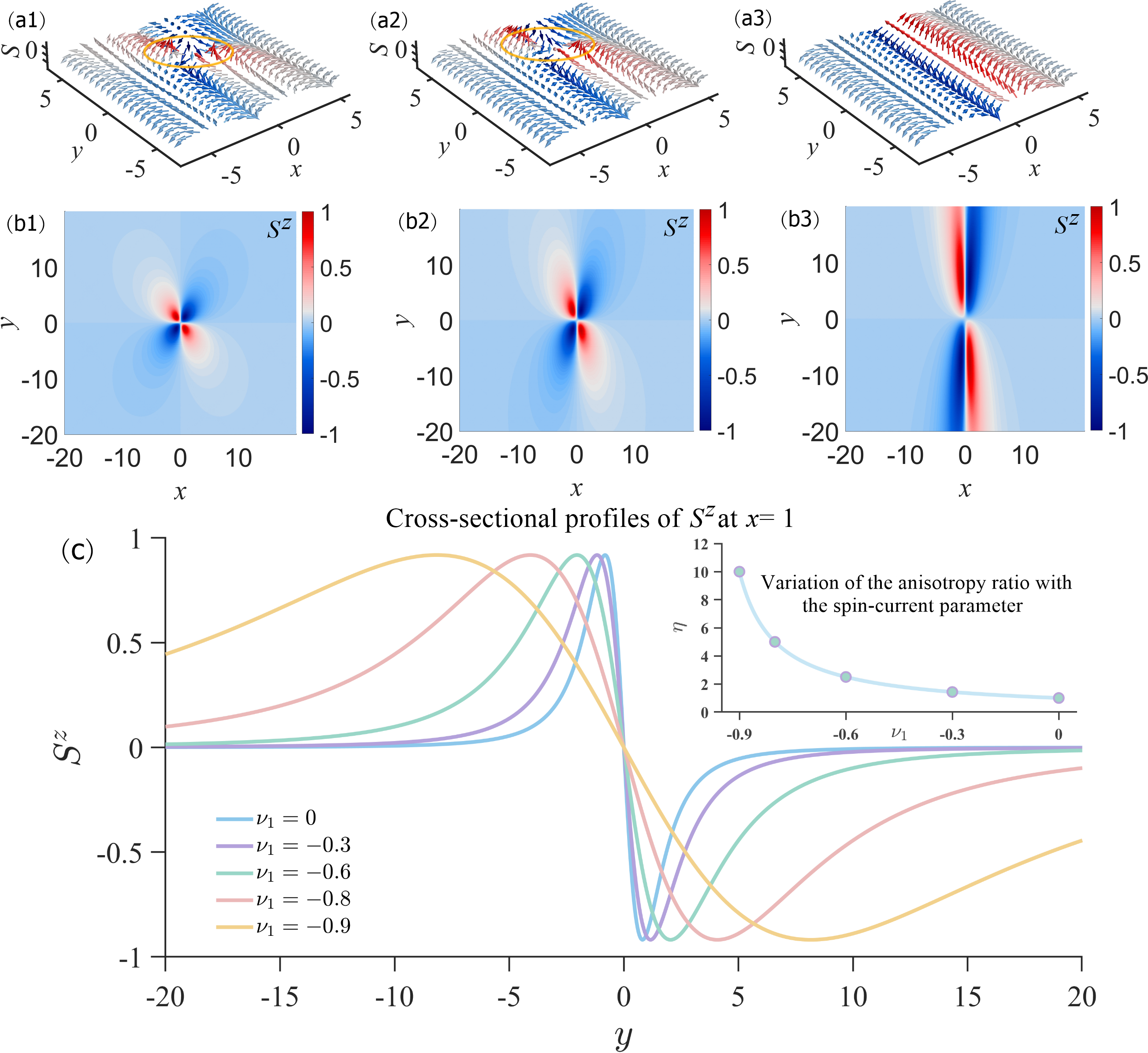}
\caption{(Color online) Spin-current-controlled structural modulation of the magnetic lump state for different values of the spin-current parameter $\nu_1$. When $t=0$, (a1)-(a3) Three-dimensional spin configurations for $\nu_1=-0.3$, $-0.6$, and $-0.9$, respectively, showing simultaneous translation and anisotropic expansion of the lump structure along the $y$-direction, and the yellow circle is clearly seen moving along the $y$ axis. (b1)-(b3) After introducing the shifted coordinate $\widetilde{y}$, corresponding density distributions of $S^z$, the center remains fixed and only the anisotropic broadening survives. (c) Cross-sectional profiles of $S^z$ at $x=1$ for different values of $\nu_1$. The progressive broadening of the profiles indicates continuous weakening of localization along the $y$-direction under spin-current modulation. }\label{fig3}
\end{figure}

Figure \ref{fig3} illustrates the spatial evolution of the magnetic lump for different values of the spin-current parameter $\nu_1$. 
For relatively small values of $\nu_1$, the excitation preserves a typical doubly localized lump profile with a well-confined core structure. 
As $\nu_1$ gradually approaches $-1$, the spatial extension along the $y$-direction becomes increasingly pronounced, while the localization along the $x$-direction remains largely unchanged. 
Consequently, the original multi-lobe lump structure undergoes continuous anisotropic deformation during the modulation process.

The three-dimensional spin configurations shown in Fig.~\ref{fig3}(a1)-(a3) demonstrate that the magnetic lump continuously expands along the $y$-direction under spin-current driving, leading to progressive deformation of the localized spin texture. 
Meanwhile, the corresponding density distributions of the out-of-plane spin component $S^z$ in Fig.~\ref{fig3}(b1)-(b3) further reveal significant redistribution of localized intensity during the evolution process. 
The initially well-confined quadrupole-like structure gradually broadens and merges along the $y$-direction, indicating continuous weakening of two-dimensional spatial confinement. To further quantify this modulation behavior, the cross-sectional profiles of $S^z$ at $x=1$ are presented in Fig.~\ref{fig3}(c). It is observed that the localization width continuously increases as $\nu_1$ approaches $-1$, while the overall profile preserves a similar localized shape during the evolution process. This behavior is consistent with the analytical localization-width relation $W_y \propto \frac{1}{|\nu_1+1|}$. Demonstrating that spin current can continuously regulate the spatial localization characteristics and morphological structure of the magnetic lump.

More importantly, the continuous deformation induced by spin current suggests that the localization dimensionality of the excitation is gradually changing during the modulation process. As the spin-current parameter approaches the critical value $\nu_1=-1$, the localization width along the $y$-direction increases continuously, implying the gradual breakdown of two-dimensional spatial confinement. Consequently, the original doubly localized lump state progressively evolves toward a lower-dimensional localized structure. This spin-current-controlled state transition will be analyzed in detail in the following subsection.

\subsection*{3.3 Lump-to-soliton state transition}

An especially interesting structural transition occurs when the spin-current parameter approaches the critical value $\nu_1=-c$. In this limit, the localization coordinate (\ref{rho}) reduces to $\rho^2=x^2+1$, which completely loses its dependence on the spatial coordinate $y$. Consequently, the excitation no longer possesses spatial confinement along the $y$-direction, indicating the breakdown of two-dimensional localization.

As demonstrated in the previous subsection, the localization width along the $y$-direction continuously increases as $\nu_1$ approaches $-c$.
This progressive weakening of spatial confinement eventually drives the system into a critical transition regime.
In particular, while the localization along the $x$-direction remains preserved, the original doubly localized lump state gradually loses its confinement along the $y$-direction, leading to a reduction of localization dimensionality.

\begin{figure}[!htbp]
\centering
\includegraphics[scale=.26]{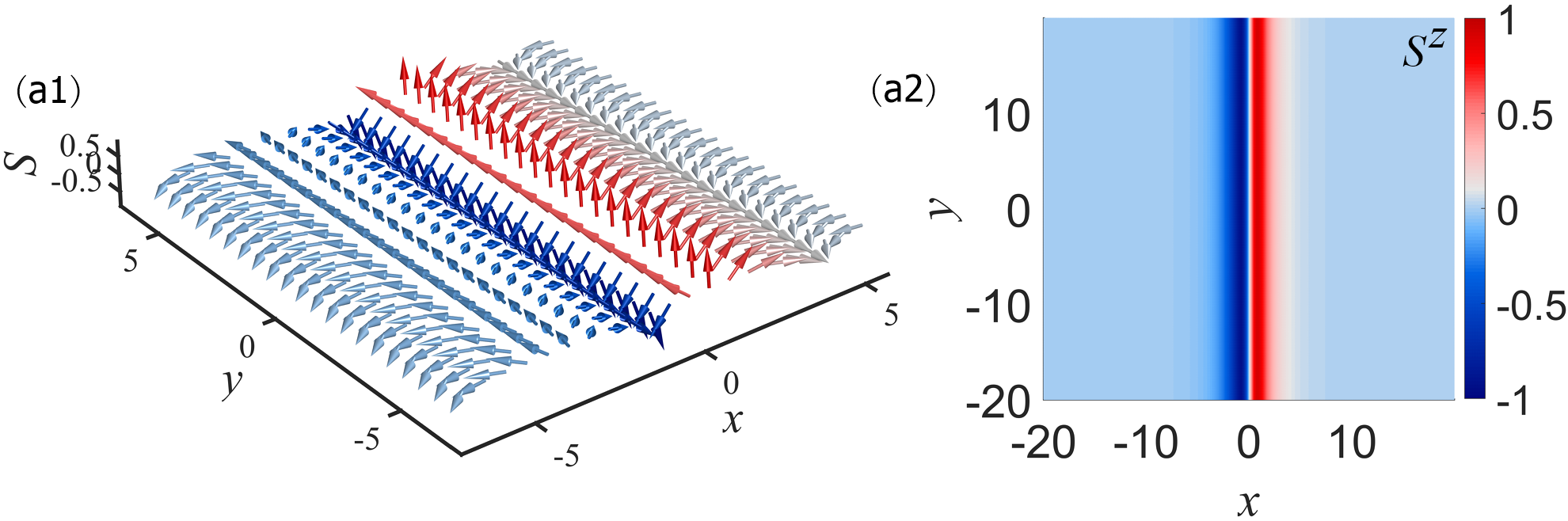}
\caption{(Color online) Line-localized limiting state at the critical point $\nu_1=-c$. 
(a1) Spatial spin-texture distribution of the resulting quasi-one-dimensional localized state. 
(a2) Density distribution of the out-of-plane spin component $S^z$, showing complete delocalization along the $y$-direction.  }\label{fig4}
\end{figure}

The corresponding structural transition is illustrated in Fig.~\ref{fig4}.
At the critical point $\nu_1=-c$, the excitation becomes completely delocalized along the $y$-direction and reduces to a soliton-like localized state possessing confinement only in the $x$-direction.
The resulting excitation therefore exhibits quasi-one-dimensional characteristics instead of the original two-dimensional lump profile.

This result demonstrates that spin current can induce a continuous dimensional transition of localized spin excitations.
Unlike conventional spin-current effects that mainly generate transport dynamics, the present mechanism directly modifies the localization dimensionality and spatial structure of the excitation itself.
The obtained lump-to-soliton transition therefore reveals a nontrivial spin-current-driven state transition behavior in nonlinear spin systems.

From a physical perspective, the spin-current parameter provides an effective external control mechanism for continuously tuning the spatial localization properties of the excitation.
By regulating the confinement behavior along the $y$-direction, the system realizes a smooth crossover from a two-dimensional doubly localized lump state to a quasi-one-dimensional soliton-like excitation.
These results provide a possible theoretical mechanism for controllable state transition and dimensional manipulation of nonlinear localized spin structures under spin-current control.

\section*{4. Universality of spin-current-induced modulation in higher-order magnetic lumps}

To further examine the generality of the spin-current-induced modulation mechanism revealed for the fundamental magnetic lump, we extend our analysis to higher-order magnetic lump solutions generated by the generalized DT. Depending on the free parameters, the second-order magnetic lump admits two representative configurations, namely the aggregated state and the separated state. These higher-order localized structures provide a natural platform for investigating whether the spin-current-controlled anisotropic deformation remains valid beyond the fundamental excitation.

\begin{figure}[!htbp]
\centering
\includegraphics[scale=.16]{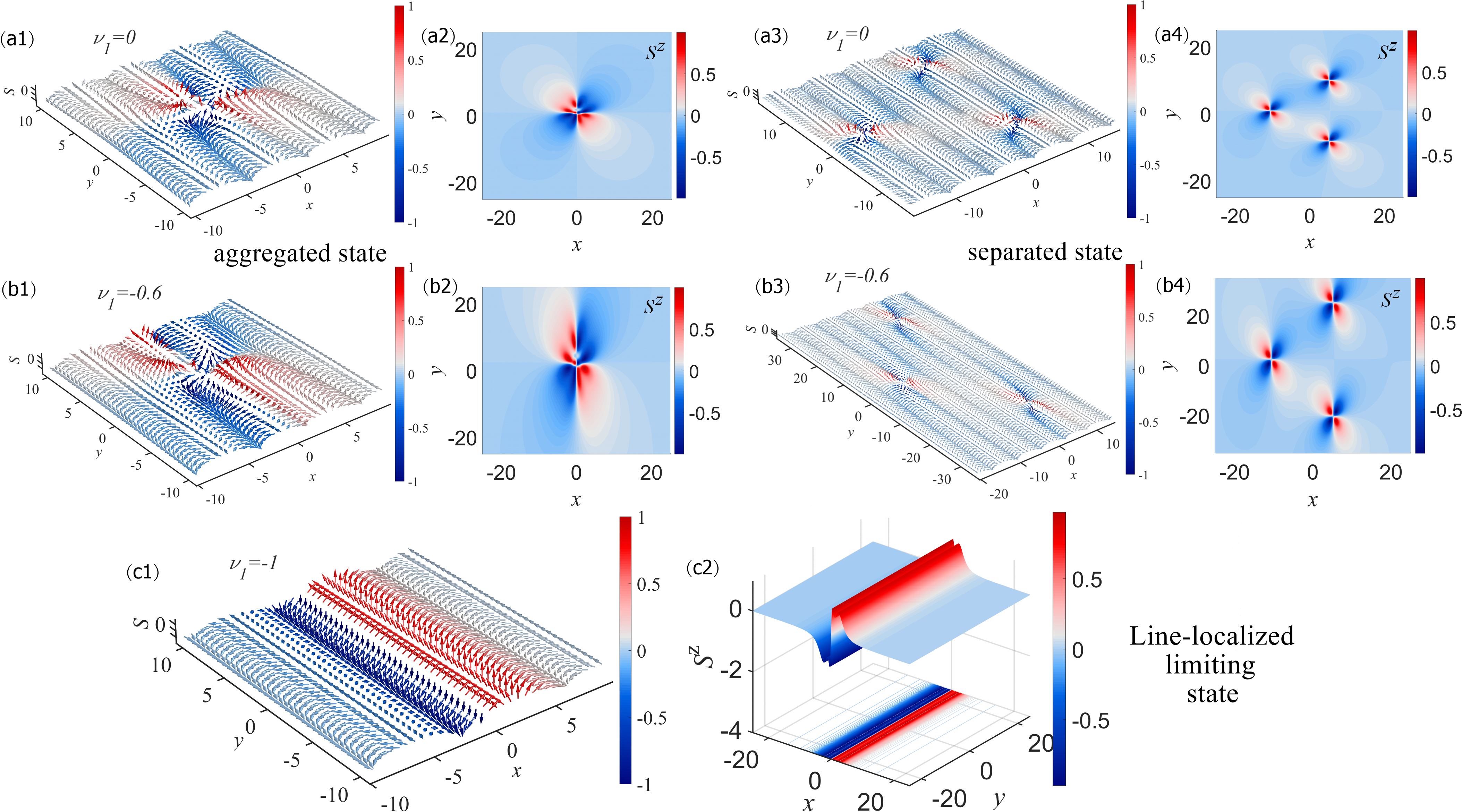}
\caption{(Color online) Universality of spin-current-induced modulation in higher-order magnetic lump structures. They still exhibit intrinsic spin precession phenomena. When $t=0$, (a1)-(a2) Aggregated-state magnetic lump and the corresponding density distribution of the out-of-plane spin component $S^{z}$ for $\nu_{1}=0$ and $e_j=d_j=0,\ (j=0,1)$. (a3)-(a4) Separated-state magnetic lump and the corresponding $S^{z}$ distribution for $\nu_{1}=0$ and $e_1=200,\ e_j=d_j=0,\ (j=0,1)$. (b1)-(b2) Aggregated-state magnetic lump and its $S^{z}$ distribution for $\nu_{1}=-0.6$ and $ e_j=d_j=0,\ (j=0,1)$, showing pronounced anisotropic broadening along the $y$ direction under spin-current modulation. (b3)-(b4) Separated-state magnetic lump and its corresponding $S^{z}$ distribution for $\nu_{1}=-0.6$ and $e_1=200,\ e_j=d_j=0,\ (j=0,1)$, exhibiting the same anisotropic deformation while preserving the interaction pattern. (c1) Higher-order magnetic lump at the critical point $\nu_{1}=-c$ and $e_j=d_j=0,\ (j=0,1)$, where the localized structure approaches a line-localized limiting state. (c2) Corresponding distribution of $S^{z}$, illustrating the disappearance of localization along the $y$ direction and the resulting quasi-one-dimensional profile.  }\label{fig5}
\end{figure}

Fig.~\ref{fig5}(a1)-(a4) display the second-order magnetic lump structures in the absence of spin-current modulation ($\nu_{1}=0$). The aggregated state consists of several strongly coupled localized cores that form a compact composite excitation, whereas the separated state contains multiple well-isolated magnetic lumps distributed in different spatial positions. Although these two configurations exhibit distinctly different spatial arrangements, both preserve the characteristic doubly localized structure inherited from the fundamental magnetic lump.

The corresponding higher-order magnetic lump structures under spin-current modulation are presented in Fig.~\ref{fig5}(b1)-(b4) for $\nu_{1}=-0.6$. Similar to the first-order case discussed in Section~3, the effective spin-current parameter induces a pronounced anisotropic deformation of both higher-order configurations. It can be observed that the localized structures continuously expand along the $y$ direction, while their localization along the $x$ direction remains nearly unchanged. Consequently, both the aggregated-state and separated-state magnetic lumps preserve their original spatial arrangement, whereas each constituent localized core undergoes an obvious transverse elongation. This indicates that spin current primarily regulates the localization geometry of each magnetic lump rather than modifying the interaction pattern among different localized structures.

As the spin-current parameter further approaches the critical value $\nu_{1}=-c$, the higher-order magnetic lump exhibits the same limiting behavior as the fundamental excitation. As shown in Fig.~\ref{fig5}(c1)-(c2), the transverse localization gradually disappears, and the localized structure approaches a line-localized soliton-like state with confinement only along the $x$ direction. Therefore, the dimensional crossover induced by spin current is preserved not only for the fundamental magnetic lump but also for higher-order localized excitations.

These results demonstrate that the spin-current-induced structural modulation established in Section~3 is not restricted to the fundamental magnetic lump. Regardless of whether the second-order solution appears as an aggregated state or a separated state, the same anisotropic broadening along the $y$ direction and the same tendency toward the line-localized limiting state are consistently preserved. This universal behavior indicates that the observed structural evolution is an intrinsic consequence of spin-current-driven localization rather than a feature associated with a particular higher-order configuration. Therefore, the spin current provides a unified physical mechanism for controlling the geometry and dimensional evolution of magnetic lump structures across different hierarchical states.

\section*{5. Conclusions and discussions }

In this work, a $(2+1)$-dimensional nonlinear spin system with an effective spin-current transport term has been investigated. Within this framework, both fundamental and higher-order magnetic lump soliton solutions have been constructed via the Darboux transformation, and the effects of spin current on the modulation of localized magnetic structures and the resulting structural transitions have been systematically analyzed.

First, in the absence of spin-current modulation, the obtained magnetic lump exhibits a spatially doubly localized stationary structure. Its spatial envelope remains invariant during temporal evolution, while the spin vector undergoes periodic precession around a fixed background. This indicates that the excitation can be regarded as a composite state consisting of a localized spatial structure and intrinsic spin precession, where the dynamical evolution manifests in the rotation of spin orientation rather than spatial propagation or translation.

Second, when the spin-current parameter is introduced, it is found that spin current does not merely affect the transport behavior of localized excitations, but directly enters the localization coordinate and fundamentally modifies the geometric structure of the magnetic lump. As the spin-current strength varies, the localized structure undergoes continuous anisotropic deformation in space, where the confinement along the $y$-direction is significantly modulated while the $x$-direction remains relatively stable. This anisotropic modulation leads to a redistribution of localized intensity and reveals a direct mechanism by which spin current controls the morphology of two-dimensional localized structures.

More importantly, the spin-current-induced modulation is found to be universal across different hierarchical lump structures. Both fundamental and higher-order magnetic lumps exhibit the same deformation mechanism, namely anisotropic stretching along the $y$-direction and a continuous transition toward a quasi-one-dimensional localized state. This universality indicates that spin current governs not only a specific solution but the entire family of lump excitations in a consistent manner.

The dimensional transition is controlled by the effective spin-current contribution, which simultaneously governs the center position and localization width of the magnetic lumps. As this contribution approaches the critical condition, the confinement along the $y$-direction is progressively weakened and eventually disappears, resulting in a continuous evolution from a two-dimensional doubly localized lump state to a quasi-one-dimensional soliton-like excitation.

It should be emphasized that the structural transition revealed in this work is fundamentally different from conventional spin-current-driven transport behavior. In most existing studies, spin current primarily induces motion or drift of magnetic structures such as domain walls, solitons, or skyrmions. In contrast, in the present model, spin current directly acts on the localization coordinate itself, leading to a reconstruction of both geometric morphology and localization dimensionality. This mechanism therefore extends the role of spin current in nonlinear magnetic systems beyond transport phenomena to structural control.

Finally, several extensions of this work are worth further investigation. In particular, the stability and dynamical response of the present two-dimensional localized structures under external perturbations, such as magnetic fields or anisotropy effects, remain open problems. Moreover, generalizing the present mechanism to higher-dimensional or non-integrable systems may reveal more complex patterns of localized structure evolution. In addition, possible experimental realizations and observable signatures of spin-current-induced dimensional transitions also deserve further study.

In summary, this work reveals a universal spin-current-controlled mechanism for the modulation and dimensional transition of magnetic lump structures across different hierarchical levels, providing a new theoretical perspective for understanding geometric control and state transitions of localized excitations in nonlinear spin systems.
\\
\\
{\bf Declaration of competing interest}

The authors declare that they have no known competing financial interests or personal relationships that could have appeared to influence the work reported in this paper.
\\
{\bf Data availability}

Data will be made available on request.
\\
{\bf Acknowledgments}

We thank the other members of the discussion group for their valuable comments.
\\
{\bf Funding}

This research was supported by Zhejiang Provincial Natural Science Foundation of China under Grant No. LQN25A050002, and the National Natural Science Foundation of China (Grant Nos. 12275213, 12247103, 12571263, 12505009, 12071042), and Beijing Natural Science Foundation (Grant No. 1242004).
\\
\\
{\bf Appendix A: The iterative generalized $(r,N-r)$-fold DT of Eq.~\eqref{hfe}}

Consider the gauge transformation
$\widetilde{\Phi}=T\Phi$,
where $T$ is a $2\times2$ Darboux matrix and
$\widetilde{\Phi}=(\widetilde{\mathcal{F}},\widetilde{\mathcal{G}})^{\rm T}$.
After the transformation, the Lax pair~(\ref{lax}) is converted into
$\widetilde{\Phi}_{x}=\widetilde{\Gamma}\widetilde{\Phi}$ and
$\widetilde{\Phi}_{t}=\lambda\widetilde{\Phi}_{y}
+\widetilde{\Xi}\widetilde{\Phi}$,
where the matrices
$\widetilde{\Gamma}$ and $\widetilde{\Xi}$
retain the same algebraic form as
$\Gamma$ and $\Xi$,
except that the original field variables
$(S^{x},S^{y},S^{z},u)$
are replaced by the transformed quantities
$(\widetilde S^{x},\widetilde S^{y},\widetilde S^{z},\widetilde u)$.
Imposing the compatibility condition gives
\begin{equation}
T_x + T\Gamma - \widetilde{\Gamma}T = 0,\qquad
T_t + T\Xi - \widetilde{\Xi}T - \lambda T_y = 0.
\label{sk}
\end{equation}

The Darboux matrix is constructed in the form
\begin{equation}
\begin{aligned}
T^{[\vartheta]}=\frac{1}{\lambda}I - M_\vartheta,\qquad
M_\vartheta = L_\vartheta \Lambda_\vartheta^{-1} L_\vartheta^{-1},
\quad (\vartheta=1,2,\cdots,N),
\\[2mm]
\Phi_\vartheta[\vartheta-1]
=
({\mathcal{F}}_\vartheta^{[\vartheta-1]},{\mathcal{G}}_\vartheta^{[\vartheta-1]})^{\rm T}
=
\left(T^{[\vartheta-1]}\cdots T^{[1]}\right)\Big|_{\lambda=\lambda_\vartheta}\Phi_\vartheta.
\end{aligned}
\label{dtz1}
\end{equation}

Here $T^{[0]}=I=\mathrm{diag}(1,1)$.
The matrices $\Lambda_\vartheta$ and $L_\vartheta$ are defined as
\begin{equation*}
\Lambda_\vartheta=
\begin{pmatrix}
\lambda_\vartheta & 0\\
0 & \lambda_\vartheta^*
\end{pmatrix},\qquad
L_\vartheta=
\begin{pmatrix}
f_\vartheta^{[\vartheta-1]} & -g_\vartheta^{*[\vartheta-1]}\\
g_\vartheta^{[\vartheta-1]} & f_\vartheta^{*[\vartheta-1]}
\end{pmatrix},
\end{equation*}
where $*$ denotes the complex conjugation.

Substituting Eqs.~(\ref{lax}), (\ref{sk}) and (\ref{dtz1}) into the Darboux framework, the first-order DT associated with Eq.~(\ref{hfe}) can be obtained as
\begin{eqnarray}
\Phi[1]=T^{[1]}\Phi[0]=(\tfrac{1}{\lambda}I-M_1)\Phi[0],\ \
\mathrm{S}[1]=\mathrm{S}[0]+2\textrm{i} M_{1x}
=\mathrm{S}[0]+2\textrm{i}(L_1\Lambda_1^{-1}L_1^{-1})_x,
\label{dt1}
\end{eqnarray}
where $T^{[1]}$ denotes the first iterative Darboux matrix evaluated at
$\lambda=\lambda_1$.
At this spectral parameter,
the corresponding eigenfunction of the Lax pair~(\ref{lax}) is
$\Phi[0]=\Phi=(\mathcal{F},\mathcal{G})^{\rm T}$,
namely,
$f_1^{[0]}=\mathcal{F}|_{\lambda=\lambda_1}$
and
$g_1^{[0]}=\mathcal{G}|_{\lambda=\lambda_1}$.
The matrices $\mathrm{S}[0]$ and $\mathrm{S}[1]$
share the same structure as the matrix $\mathrm{S}$ appearing in the Lax pair,
with the field variables
$S^{x}$, $S^{y}$, $S^{z}$ and $u$
replaced by
$S^{x}[j]$, $S^{y}[j]$, $S^{z}[j]$ and $u[j]$
after each iteration
($j=\vartheta-1$, $\vartheta=1,2,\ldots,N$).
The case $j=0$ corresponds to the prescribed seed solution.

Applying the above procedure successively, let
$\Phi_\vartheta$
($\vartheta=1,2,\ldots,N$)
be $N$ linearly independent eigenfunctions of the Lax pair~(\ref{lax}),
associated with the spectral parameters
$\lambda_\vartheta$.
The iterative $N$-fold Darboux transformation is then expressed as
\begin{equation}
\begin{array}{ll}
\Phi[N]=T^{[N]}\Phi[N-1]
=T^{[N]}T^{[N-1]}\cdots T^{[1]}\Phi[0],\quad
\mathrm{S}[N]
=\mathrm{S}[N-1]+2\textrm{i}M_{Nx}
=\mathrm{S}[0]+2\textrm{i}\sum\limits_{\vartheta=1}^{N}M_{\vartheta x},
\end{array}
\label{dt2}
\end{equation}
which provides the recursive construction of higher-order exact solutions.

If the Taylor expansion is not introduced for the eigenfunction $\Phi$ in the gauge transformation and the iteration number $N$ coincides with the number of distinct spectral parameters, the iterative DT generates multi-soliton or breather solutions on different backgrounds. However, when the iteration number is $N$ while the spectral parameter $\lambda$ assumes only $r\ (1\leq r < N)$ distinct values, the standard iterative DT can no longer produce new solutions of Eq.~(\ref{hfe}). To overcome this difficulty, the eigenfunction $\Phi$ is expanded in a Taylor series, namely,
$\Phi(w_\iota+\varepsilon)=\Phi_\iota^{[0]}+\Phi_\iota^{[1]}\varepsilon+\Phi_\iota^{[2]}\varepsilon^2+
\Phi_\iota^{[3]}\varepsilon^3+\cdots$, where
$\Phi_\iota^{[\kappa]}=\frac{1}{\kappa!}\frac{\partial^\kappa}{\partial w^\kappa}\Phi_\iota(w)|_{w=w_\iota}\ (\iota=1,2,\ldots,r,\ \kappa=0,1,2,\ldots)$,
$\varepsilon$ is an infinitesimal parameter, and
$w_\iota=\frac{1}{\lambda_\iota}$.

For the special case $r=1$, where only a single spectral parameter is involved, the generalized $(1,N-1)$-fold iterative DT is given by
\begin{equation}
\begin{array}{ll}
\Phi[N]=T_1^{[N]}\Phi[N-1]=T_1^{[N]}T_1^{[N-1]}...T_1^{[1]}\Phi_1^{[0]},\quad
\mathrm{S}[N]=\mathrm{S}[N-1]+2\textrm{i} M_1[N]_{x}=\mathrm{S}[0]+2\textrm{i} \sum\limits_{\vartheta=1}^NM_1[\vartheta]_{x},
\end{array}\label{r1}
\end{equation}
where $\Phi_1^{[0]}=(\mathcal{F}_1^{[0]},\mathcal{G}_1^{[0]})^{\rm T}$ is the first term in the Taylor expansion of $\Phi(w_\iota+\varepsilon)$ evaluated at $\lambda=\lambda_1$, and
\begin{equation}
\begin{array}{cc}
T_1^{[\vartheta]}=\frac{1}{\lambda}I-M_1[\vartheta],\quad M_1[\vartheta]=L_1[\vartheta]\Lambda_1^{-1}L_1[\vartheta]^{-1}, \quad (\vartheta=1,2,...,N),\vspace{0.1in}\\
\Lambda_1 = \left(
\begin{array}{cc}
\lambda_1 & 0 \vspace{0.05in}\\ 0 & \lambda_1^*
\end{array}
\right) ,\quad
M_1[\vartheta]=\left(
\begin{array}{cc}
f_1^{[\vartheta-1]} & -g^{*[\vartheta-1]}_1 \\ g_1^{[\vartheta-1]} & f^{*[\vartheta-1]}_1\\
\end{array}
\right),
\end{array} \label{r2}
\end{equation}
The corresponding eigenfunction of the Lax pair~(\ref{lax}) is then obtained through the limiting process
\begin{equation}
\begin{array}{cc}
\Phi_1[\vartheta-1]=\left(
\begin{array}{cc}
\mathcal{F}_{1}^{[\vartheta-1]} & \mathcal{G}_{1}^{[\vartheta-1]}
\end{array}
\right)^{\mathrm{T}}=\lim\limits_{\varepsilon \rightarrow 0}\dfrac{T_1^{[\vartheta-1]}T_1^{[\vartheta-2]}\cdots T_1^{[1]}|_{w=w_1+\varepsilon}\Phi(w_1+
\varepsilon)}{\varepsilon^{\vartheta-1}}.
\end{array} \nonumber
\end{equation}

The above construction can be naturally extended to the general case $1<r<N$. The corresponding fundamental solution of the Lax pair~(\ref{lax}) is recursively determined by
\begin{equation}
\begin{array}{ll}
\Phi_\iota[p]=\left(
\begin{array}{cc}
\mathcal{F}_{\iota}^{[p]} & \mathcal{G}_{\iota}^{[p]}
\end{array}
\right)^{\mathrm{T}}=\lim\limits_{\varepsilon \rightarrow 0}\dfrac{T_\iota^{[p]}T_\iota^{[p-1]}\cdots T_\iota^{[1]}|_{w=w_\iota+\varepsilon}\Phi_\iota
(w_\iota+\varepsilon)}{\varepsilon^p}\vspace{0.1in}\\ \qquad\
=\Phi_\iota^{[0]}+(T_\iota^{[p]}+T_\iota^{[p-1]}+\cdots+T_\iota^{[1]})|_{w=w_\iota}\Phi_\iota^{[1]}
+\cdots+(T_\iota^{[p]}T_\iota^{[p-1]}\cdots T_\iota^{[1]})|_{w=w_\iota}\Phi_\iota^{[p]},
\end{array} \label{r3}
\end{equation}
where $T_\iota^{[p]}\ (\iota=1,2,\ldots,r,\ p=1,2,\ldots)$ is obtained by replacing $(\mathcal{F}_{\vartheta}^{[\vartheta-1]},\mathcal{G}_{\vartheta}^{[\vartheta-1]})$ in Eq.~(\ref{dtz1}) with $(\mathcal{F}_{\iota}^{[p-1]},\mathcal{G}_{\iota}^{[p-1]})$, while the corresponding spectral parameter is replaced by $\lambda_\iota$. In Eq.~(\ref{r3}), the positive integer $p$ satisfies the constraint $1\leq p\leq \nu_\iota+1$. For each spectral parameter $\lambda_\iota$ satisfying the condition $N=r+\sum\limits_{\iota=1}^{r}\delta_\iota$, the integer $\delta_\iota$ denotes the highest derivative order retained in the Taylor expansion of the eigenfunction of the Lax pair~(\ref{lax}). Consequently, the potential functions are transformed according to
\begin{equation}
\label{xjst}
\begin{array}{ll}
\widetilde{S}^{x}[N]=S^{x}[\kappa]+\mathrm{i}K_1\Bigl[(g_{\iota}^{[\kappa]})^{*}\bigl((f_{\iota}^{[\kappa]})^{2}-
(g_{\iota}^{[\kappa]})^{2}\bigr)(f_{\iota,x}^{[\kappa]})^{*}-(f_{\iota}^{[\kappa]})^{*}\bigl((f_{\iota}^{[\kappa]})^{2}-
(g_{\iota}^{[\kappa]})^{2}\bigr)(g_{\iota,x}^{[\kappa]})^{*}
\\[4mm] \qquad \qquad
+\Bigl[\bigl((f_{\iota}^{[\kappa]})^{*}\bigr)^{2}-\bigl((g_{\iota}^{[\kappa]})^{*}\bigr)^{2}\Bigr]W_x\Bigr],
\\[4mm]
\widetilde{S}^{y}[N]=S^{y}[\kappa]+K_1\Bigl[-(g_{\iota}^{[\kappa]})^{*}\bigl((f_{\iota}^{[\kappa]})^{2}+
(g_{\iota}^{[\kappa]})^{2}\bigr)(f_{\iota,x}^{[\kappa]})^{*}+(f_{\iota}^{[\kappa]})^{*}\bigl((f_{\iota}^{[\kappa]})^{2}
+(g_{\iota}^{[\kappa]})^{2}\bigr)(g_{\iota,x}^{[\kappa]})^{*}
\\[2mm]\qquad \qquad
+\Bigl[\bigl((f_{\iota}^{[\kappa]})^{*}\bigr)^{2}+\bigl((g_{\iota}^{[\kappa]})^{*}\bigr)^{2}\Bigr]W_x
\Bigr],
\\[4mm]
\widetilde{S}^{z}[N]=S^{z}[\kappa]-2\mathrm{i}K_1\Bigl[f_{\iota}^{[\kappa]}g_{\iota}^{[\kappa]}(g_{\iota}^{[\kappa]})^{*}
(f_{\iota,x}^{[\kappa]})^{*}-f_{\iota}^{[\kappa]}g_{\iota}^{[\kappa]}(f_{\iota}^{[\kappa]})^{*}
(g_{\iota,x}^{[\kappa]})^{*}-f_{\iota}^{[\kappa]}(f_{\iota}^{[\kappa]})^{*}(g_{\iota}^{[\kappa]})^{*}
g_{\iota,x}^{[\kappa]}
\\[4mm]\qquad \qquad
+f_{\iota}^{[\kappa]}(f_{\iota}^{[\kappa]})^{*}(g_{\iota}^{[\kappa]})^{*}f_{\iota,x}^{[\kappa]}\Bigr],
\\[4mm]
\widetilde{u}^{[N]}=\dfrac{{S}^{z}[\kappa]\,{u}{[\kappa]}}{\widetilde{S}^{z}[N]}+
\dfrac{\mathcal{P}_x+\mathcal{P}_t}{|\lambda_{\iota}|^{2}K_2^{2}\widetilde{S}^{z}[N]}.
\end{array}
\end{equation}
with
\begin{equation*}
\label{KP}
\begin{array}{ll}
K_1=\frac{\lambda_{\iota}-\lambda_{\iota}^{*}}{|\lambda_{\iota}|^2K_2^2}, \quad
K_2=|f_{\iota}^{[\kappa]}|^2+|g_{\iota}^{[\kappa]}|^2, \quad
W_x=g_{\iota}^{[\kappa]}f_{\iota,x}^{[\kappa]}-f_{\iota}^{[\kappa]}g_{\iota,x}^{[\kappa]},
\\[2mm]
\mathcal P_x=\Bigl(\nu_1S^{z}[\kappa]-\nu_1\widetilde S^{z}[N]-S^{y}[\kappa]S_y^{x}[\kappa]+\widetilde S^{y}[N]\widetilde S_y^{x}[N]
\\[2mm] \quad \quad
+S^{x}[\kappa]S_y^{y}[\kappa]-\widetilde S^{x}[N]\widetilde S_y^{y}[N]\Bigr)|\lambda_{\iota}|^2K_2^2,
\\[2mm]
\mathcal P_t=-2\mathrm{i}\lambda_{\iota}^{*}\Bigl[(g_{\iota}^{[\kappa]})^{*}f_{\iota}^{[\kappa]}
g_{\iota,t}^{[\kappa]}-(g_{\iota}^{[\kappa]})^{*}g_{\iota}^{[\kappa]}f_{\iota,t}^{[\kappa]}+
f_{\iota}^{[\kappa]}g_{\iota}^{[\kappa]}(g_{\iota,t}^{[\kappa]})^{*}\Bigr](f_{\iota}^{[\kappa]})^{*}
+2\mathrm{i}\lambda_{\iota}^{*}f_{\iota}^{[\kappa]}g_{\iota}^{[\kappa]}(g_{\iota}^{[\kappa]})^{*}
(f_{\iota,t}^{[\kappa]})^{*}
\\[2mm] \quad \quad
-2\mathrm{i}\lambda_{\iota}\Bigl[\bigl(-f_{\iota}^{[\kappa]}g_{\iota,t}^{[\kappa]}+g_{\iota}^{[\kappa]}
f_{\iota,t}^{[\kappa]}\bigr)(g_{\iota}^{[\kappa]})^{*}-f_{\iota}^{[\kappa]}g_{\iota}^{[\kappa]}
(g_{\iota,t}^{[\kappa]})^{*}\Bigr](f_{\iota}^{[\kappa]})^{*}
-2\mathrm{i}\lambda_{\iota}f_{\iota}^{[\kappa]}g_{\iota}^{[\kappa]}(g_{\iota}^{[\kappa]})^{*}
(f_{\iota,t}^{[\kappa]})^{*}.
\end{array}
\end{equation*}
In the above expressions, $\widetilde{S^x}[N]$, $\widetilde{S^y}[N]$, $\widetilde{S^z}[N]$, and $\widetilde{u}[N]$ denote the transformed solutions obtained after the $N$-th iteration, where $\kappa=N-1$, $f_\iota^{[\kappa]}=\mathcal{F}_{\iota}^{[p-1]}$, and $g_\iota^{[\kappa]}=\mathcal{G}_{\iota}^{[p-1]}$. As defined previously, when $\kappa=0$, $S^x[0]$, $S^y[0]$, $S^z[0]$, and $u[0]$ correspond to the seed solutions. For $\kappa>0$, the iterated fields satisfy $S^j[\kappa]=\widetilde{S^j}[N-1]$ and $u[\kappa]=\widetilde{u}[N-1]$, where $j=x,\ y,\ z$, $\kappa=1,2,\ldots$, and $N=2,3,\ldots$. We are now in a position to state the iterative generalized $(r,N-r)$-fold DT theorem.

{\bf Theorem 1.} Let $\Phi_\iota=(\mathcal{F}_\iota,\ \mathcal{G}_\iota)^{\rm T}$ be $r$ fundamental solutions of the Lax pair~(\ref{lax}) associated with the spectral parameters $\lambda_\iota\ (\iota=1,2,\ldots,r)$. Let $S^x[0]$, $S^y[0]$, $S^z[0]$, and $u[0]$ be the seed solutions of Eq.~(\ref{hfe}). Then, the iterative generalized $(r,N-r)$-fold DT is given by
\begin{equation}
\begin{array}{ll}
\Phi[N]=T\Phi[0], \quad
\mathrm{S}[N]=\mathrm{S}[N-1]+2\textrm{i} M_r[p]_x=\mathrm{S}[0]+2\textrm{i} \sum\limits_{\iota=1}^r \sum\limits_{p=1}^{\delta_\iota} M_\iota[p]_x,
\end{array}\label{dt3}
\end{equation}
where $T=Q_r Q_{r-1}\cdots Q_0$, $Q_\iota=T_\iota^{[\nu_\iota+1]}\cdots T_\iota^{[1]}\ (\iota=1,2,\ldots,r)$, $\Phi_\iota[0]=(\mathcal{F}_\iota^{[0]},\ \mathcal{G}_\iota^{[0]})^{\rm{T}}=(Q_{\iota-1}\cdots Q_0)|_{\frac{1}{\lambda}=\frac{1}{\lambda_\iota}}\Phi_\iota^{[0]}$, and $Q_0=I$. Here,
\begin{equation}\small
T_\iota^{[p]}=\dfrac{1}{\lambda}I-M_\iota[p]=\left(
\begin{array}{cc}
\frac{1}{\lambda}-M_{11\iota}[p] & -M_{12\iota}[p] \vspace{0.05in}\\ -M_{21\iota}[p] & \frac{1}{\lambda}-M_{22\iota}[p] \\
\end{array}\nonumber
\right),
\end{equation}
where
\begin{equation}
\begin{array}{ll}
M_{11\iota}[p]=\dfrac{\lambda_\iota^*|\mathcal{F}_{\iota}^{[p-1]}|^2+\lambda_\iota|\mathcal{G}_{\iota}^{[p-1]}|^2}
{\left(|\mathcal{F}_{\iota}^{[p-1]}|^2+|\mathcal{G}_{\iota}^{[p-1]}|^2\right)|\lambda_\iota|^2},
\quad
M_{12\iota}[p]=\dfrac{(\lambda_\iota^*-\lambda_\iota)\mathcal{F}_{\iota}^{[p-1]}\mathcal{G}_{\iota}^{[p-1]*}}
{\left(|\mathcal{F}_{\iota}^{[p-1]}|^2+|\mathcal{G}_{\iota}^{[p-1]}|^2\right)|\lambda_\iota|^2},
\vspace{0.1in}\\
M_{21\iota}[p]=\dfrac{(\lambda_\iota^*-\lambda_\iota)\mathcal{F}_{\iota}^{[p-1]*}\mathcal{G}_{\iota}^{[p-1]}}
{\left(|\mathcal{F}_{\iota}^{[p-1]}|^2+|\mathcal{G}_{\iota}^{[p-1]}|^2\right)|\lambda_\iota|^2},
\quad
M_{22\iota}[p]=\dfrac{\lambda_\iota|\mathcal{F}_{\iota}^{[p-1]}|^2+\lambda_\iota^*|\mathcal{G}_{\iota}^{[p-1]}|^2}
{\left(|\mathcal{F}_{\iota}^{[p-1]}|^2+|\mathcal{G}_{\iota}^{[p-1]}|^2\right)|\lambda_\iota|^2},
\end{array}
\nonumber
\end{equation}
where $M_\iota[p]_x\ (p=1,2,\ldots)$ denotes the derivative of $M_\iota[p]$ with respect to $x$. For $p>1$, the vector $(\mathcal{F}_\iota^{[p-1]},\ \mathcal{G}_\iota^{[p-1]})^{\rm T}$ is determined recursively from Eq.~(\ref{r3}).

\end{document}